Topic 14. Retrofit and optimal operation of the building energy systems

# Performances of Low Temperature Radiant Heating Systems

Milorad Bojić[1*,] Dragan Cvetković[1], Jasmina Skerlić[1], Danijela Nikolić[1], Harry Boyer[2]

[1]Faculty of Engineering, University of Kragujevac, Serbia
[2]Department PIMENT Lab., University of Réunion Island, France

[*]Corresponding email: bojic@kg.ac.rs

**SUMMARY**

Low temperature heating panel systems offer distinctive advantages in terms of thermal comfort and energy consumption, allowing work with low exergy sources. The purpose of this paper is to compare floor, wall, ceiling, and floor-ceiling panel heating systems in terms of energy, exergy and CO2 emissions.

Simulation results for each of the analyzed panel system are given by its energy (the consumption of gas for heating, electricity for pumps and primary energy) and exergy consumption, the price of heating, and its carbon dioxide emission. Then, the values of the air temperatures of rooms are investigated and that of the surrounding walls and floors.

It is found that the floor-ceiling heating system has the lowest energy, exergy, CO2 emissions, operating costs, and uses boiler of the lowest power. The worst system by all these parameters is the classical ceiling heating.

**INTRODUCTION**

In Europe today, low-temperature panel heating and cooling systems for residential buildings are increasingly used. According to some studies, this figure exceeds 50% (Kilkis et al. 1994). According to some studies, energy saving by panel systems is more than 30% than that by the ordinary heating systems (Stetiu 1999; Yost et al. 1995).

The low temperature radiant systems are very complex because they involve different mechanisms of heat transfer: heat conduction through the walls, heat convection between the heating panel and the indoor air, heat radiation between the heating panel and the surrounding areas, and the heat conduction between the floor and the ground. The main essence of the low-temperature air systems is to provide adequate thermal comfort at significantly lower temperatures.

A large number of studies is devoted to laboratory tests of panel systems in terms of heat transfer research and development of new calculation methods. However, in terms of heat transfer modelling, there are several analytical studies on thermal characteristics of panel systems. The earliest model by Kollmar and Liese (1957) showed that heat loss is mainly from the upper surface of the floor panel board. Zhang and Pate (1987) developed a two-dimensional finite element method for the ceiling panel heating, which is used to model low-temperature heating systems. Kilkis et al. (1995) developed the so-called stationary composite model for modelling of radiant systems for heating and cooling. After that, Kilkis and Coley

(1995) used this model and developed the software for design of floor heating and cooling systems. Maloney et al. (1988) developed a model for radiant heating panels for BLAST software. Strand and Pedersen (1997, 2002) used the method of transfer by heat conduction and developed a model for radiant heating and cooling within the EnergyPlus. Miriel et al. (1999) used TRNSYS software for modelling the ceiling panels for heating and cooling. Laouadi (2004) developed one-dimensional numerical software with two-dimensional analytical model of floor/ceiling heating systems. Compared to earlier works, based on one-dimensional numerical models, this work provides better prediction of the temperature between the concrete and the pipes. This leads to more accurate calculation of other necessary parameters (thermal comfort, capacity boilers, heating control, the minimum and maximum floor temperature). So far, no published papers are found devoted to the comparison of larger number of the panel heating systems from different points of view: energy, exergy, and environmental protection.

This paper presents research with an objective to compare the heating effects of several panel heating systems: floor, wall, ceiling, and floor-ceiling. These low-temperature heating systems are simulated for the same residential house by using simulation models of EnergyPlus. The obtained parameters of energy and exergy consumption and that of $CO_2$ emission are mutually compared for the simulated cases and different conclusions are obtained.

**MATHEMATICAL MODEL**

**Building description**
The analyzed building is a residential family house shown in Figure 1. The house is designed for one family and has a living area of 190 m$^2$. The envelope of the house is made of 190 mm porous brick, 50 mm thermal insulating layer and 20 mm lime mortar. The U-value is 0,57W/ (m$^2$K). The windows are double glazed with U-value of 2.72 W/ (m$^2$K). The overall ratio of glass to the exterior walls is 7.32%, where the total area of exterior walls is 264 m$^2$ and area of windows 19 m$^2$.

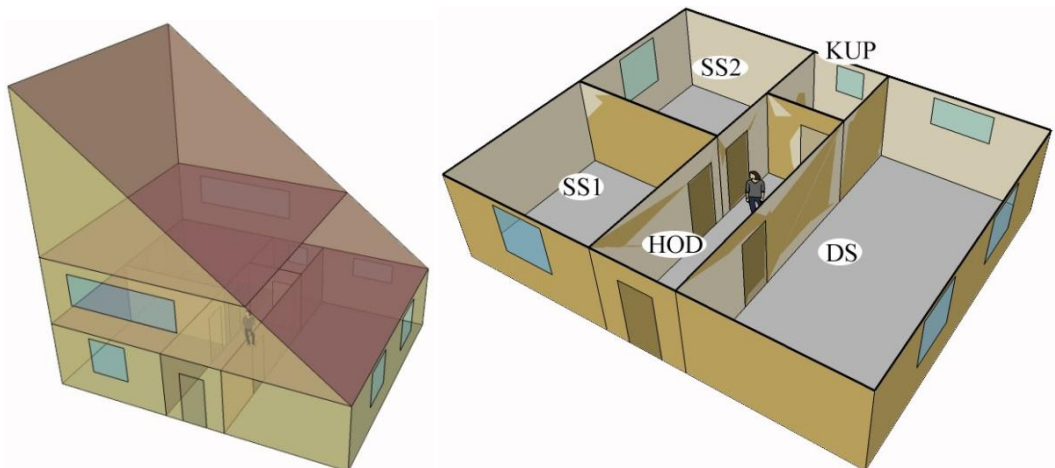

*Figure 1 Analyzed building*

The analyzed house is located in Kragujevac, Serbia. The elevation of Kragujevac is 209 m, and its latitude and longitude are 44°N and 20°55E. The city has a continental temperate climate with four different seasons (summer, autumn, winter, and spring). As part of the EnergyPlus, weather file used as an epw file generated by the Meteonorm *(World weather, 2011)*. The heating season runs in Kragujevac from 15 October to 15 April (Bogner 2002).

**Description of panel systems**
The panel heaters may be the floor heating panels, wall heating panels, ceiling heating panels, and floor-ceiling heating panels. The floor heating panel has the total surface area of 190 m$^2$. The wall heating panel is located at the external wall. Its total surface area is 210 m$^2$. The ceiling heating panel is located at the ceiling of the first and second storey of the house. Its total surface area is 190 m$^2$. The floor-ceiling heating panel operates as a ceiling heating of the lower story, and as a floor heating of the upper story. Its total surface area is 95 m$^2$.

The main component of the heating panels is the pipe where the hot water flows. The hot water inlet temperature has the same value of 37$^o$C for all heating systems.

For all heating panels, the classic non-condensing boilers are used to generate heat by using natural gas. The water circulation pump uses electricity to operate. This is taken into account to calculate the primary energy consumption.

Table 2 gives the length of the pipes in the panel heating systems. These lengths directly depend of the architecture of the house.

*Table 2 The pipe length in the panels*

| The panel heating systems | The pipe length, m |
|---|---|
| Floor | 1267 |
| Wall | 1007 |
| Ceiling | 1068 |
| Floor-ceiling | 634 |

Four systems are investigated. The first heating system represents the floor heating. The second heating system represents the wall panel heating. The third heating system represents the ceiling heating. The fourth heating system represents the floor-ceiling heating.

**Primary energy consumption of heating system**
The primary energy consumption per heating season of the analyzed house is calculated by using the following equation:

$$E_{sys} = E_{ng} + R\, E_{el} \tag{1}$$

Here, $E_{ng}$ stands for the consumption of natural gas per heating season, $E_{el}$ stands for the consumption of electricity per heating season and R stands for the primary energy consumption coefficient. This coefficient defined as the ratio of the total input energy of energy resources (hydro, coal, oil and natural gas) and the finally produced electric energy. Its value for the Serbian energy mix for electrical energy production is R = 3.61 (The energy balance, 2011).

**Consumed exergy for heating**
The consumed exergy per heating season of analyzed house is calculated by using the following equation:

$$E_{X_T} = \sum_{i=1}^{n} E_{Xi} = \sum_{i=1}^{n=20} \left(1 - \frac{T_o}{\left(\frac{T_{in_i} + T_{ret_i}}{2}\right)}\right) E_{ng_i} \tag{2}$$

Here, $n$ stands number of heating rooms, $T_o$ stands for reference temperature, $T_{ini}$ and $T_{reti}$ stands for inlet and return temperatures of heating emission systems for observed room.

**Carbon dioxide emission**

The total carbon dioxide emission of heating system during a system operation is calculated by using the following equation:

$$S_{sys} = g_{ng} E_{ng} + g_{el} E_{el} \tag{3}$$

Here, $g_{ng}$ stands for specific carbon dioxide emission factor of natural gas (kg/GJ), $g_{el}$ stands for specific carbon dioxide emission factor of electrical energy (kg/GJ). The stands for emission factors $g_{ng}$ and $g_{el}$ are 56.1 and 206.53 respectively (The energy balance, 2011).

**Operating cost**

The totally operating costs to run a system are calculated by using the following equation:

$$C_{TOT} = f_{ng} E_{ng} + k\, m_1 f_{el} E_{el} \tag{4}$$

here, $f_{ng}$ stands for the specific cost of consumption of natural gas with energy value of 33338 kJ/m$^3$ (in €/GJ), and $f_{el}$ stands for the specific cost of consumption of electrical energy (in €/GJ), $k$ stands for the coefficient of correction quantity of consumed gas k=1.068 and $m_1$ stands fixed monthly cost for meter reading (Interklima, method of calculation, 2011). In this equation we don't include fixed monthly cost for meter reading of electricity energy because we have more electricity consumer which is a much larger part. The cost factors are given in Table 1.

*Table 1. The price of energy in Serbia in May 2011 (Interklima, the price of natural gas, 2011; Interklima, method of calculation, 2011)*

| Final energy | Class of consumption | Price |
|---|---|---|
| **Electric energy** | $f_{el}$ for green tariff (<350 kWh)[1] | 0.059 €/kWh |
| | $f_{el}$ for blue tariff (351-1600 kWh) | 0.089 €/kWh |
| | $f_{el}$ for red tariff (>1601 kWh) | 0.177 €/kWh |
| **Natural gas** | $f_{ng}$ | 0.41 €/m$^3$ |
| | fixed monthly cost for meter reading $m_1$ | 0.012 €/month |

[1]Monthly electricity consumption per house

# RESULTS AND DISCUSSION

## Results

Simulation results for each of the analyzed panel system are expressed through its energy and exergy consumption, the price of heating (calculated according to current tariffs in Serbia for 2011), and its carbon dioxide emission. Figure 2 shows the consumption of gas, electricity for pumps, and primary energy during heating season. Figure 3 shows the price of heating.

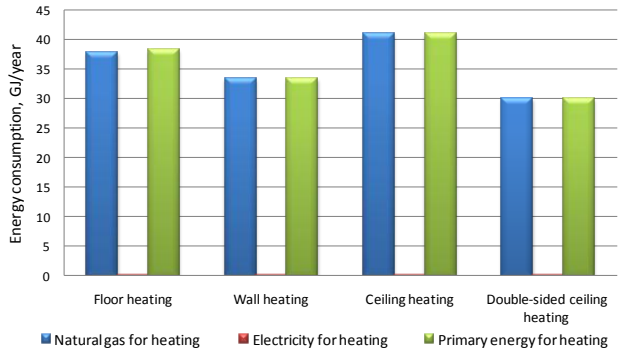

Figure 2. Energy consumption per heating season

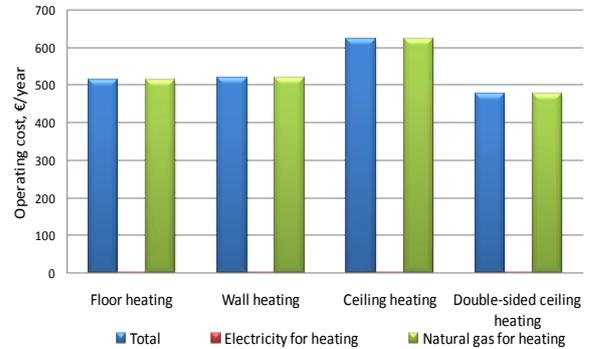

Figure 3. Heating costs per heating season

Figure 4 shows the value of the nominal power of the boiler for all four cases. Figure 5 shows the CO2 emissions per heating season. Figure 6 shows the exergy consumption per heating season.

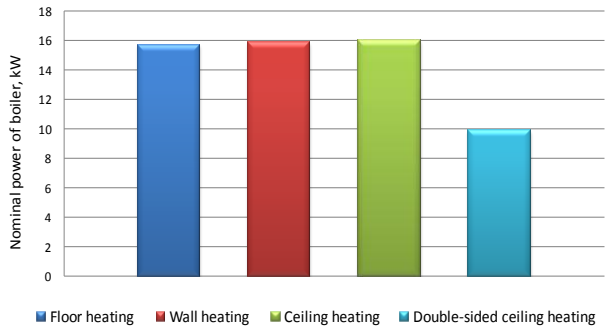

Figure 4. Nominal power of boilers

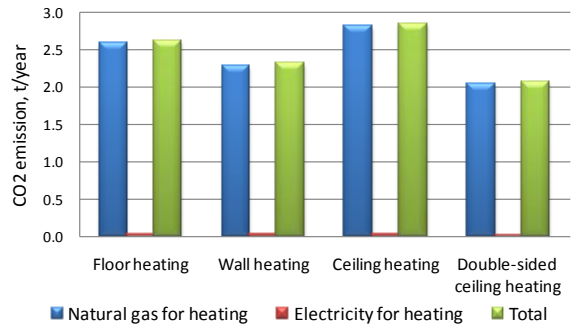

Figure 5. $CO_2$ emission per heating season

Figure 7 shows the obtained values of the mean air temperature of rooms, and the desired temperature of rooms in January.

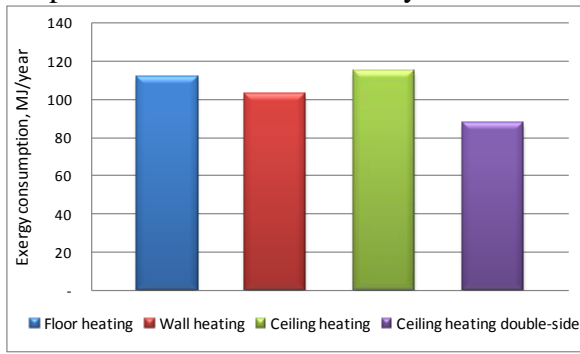

Figure 6. Exergy for January

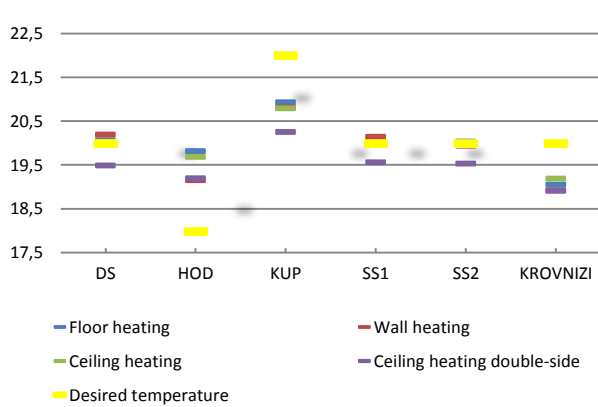

Figure 7. The mean and desired temperatures of rooms in January

Figure 8 shows the mean temperatures of the surrounding walls and floors in January.

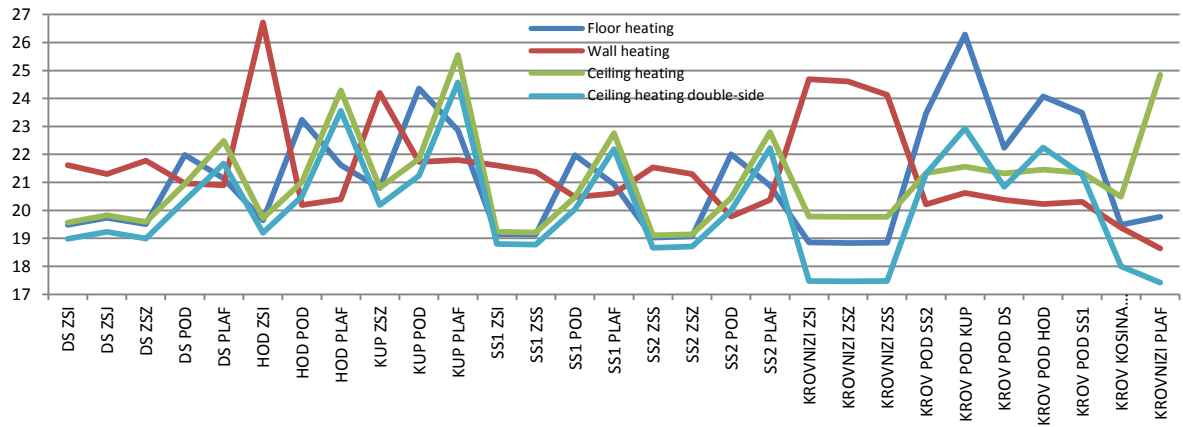

*Figure 8 The mean temperatures of surroundings walls, floors and ceilings during January*

**Discussion**

It is found that the system 4 has the lowest primary energy consumption (29.9 GJ/year) and the system 3 has the highest primary energy consumption (41.1 GJ/ year). Their difference is about 27%. The system 4 has the primary energy consumption lower by 10% than that of system 2, and lower by 22% than that of system 1. The electricity consumption slightly influences the overall energy required. The lowest electricity consumption is with system 4 (0.08 GJ/ year), and highest electricity consumption with system 3 (0.13 GJ/ year). Figure 3 shows that the largest heating cost is that with heating system 3 (624 €), and the lowest with heating system 1 (475 €). The use of system 1 instead of system 3 yields a total financial savings of 148 € per heating season. The minimum nominal power of the boiler is required for system 4 (9.89 kW), and the highest for system 3 (15.99 kW). The nominal power outputs by other two systems are similar as by system 3. The $CO_2$ emissions are the highest by system 3 (2.85 $tCO_2$), and the lowest by system 4 (2.07 $tCO_2$). Figure 6 shows the consumed exergy in January when outdoor temperatures are the lowest. Then, the exergy is the smallest by system 4 (87 MJ), and highest by system 3 (115 MJ).

To check proper operation of all four systems, Figure 7 shows the mean internal air temperature in relation to the temperature set by thermostats. It is found that for all heating systems, the mean internal air temperatures do not significantly deviate from the design temperature. The largest deviation occurs with heating system 4 in room SS3. The value of this deviation is about 2 K, so that the mean temperature in this room is 18.0°C instead of the desired 20°C. To monitor the appearance of unsolicited overheating and too cold walls, Figure 8 shows the mean temperatures of indoor surfaces of the exterior walls. All temperatures are in the range between 17.5 and 27 °C, but the temperature jump at Figure 8 is a phenomenon characteristic for the wall where the low temperature panel is integrated.

**CONCLUSIONS**

This paper analyzes the four low-temperature radiant panel systems: floor heating, wall heating, ceiling heating, and ceiling-floor heating. The classic (no condensing) natural gas boiler is used as a source of energy. The boiler is connected to a circulation pump with constant flow. The family house for the city of Kragujevac is analyzed with architecture developed within the project "Development of a net-zero-energy house." The operating of each heating system is evaluated through the consumption of primary energy, exergy, operating cost, and emissions of pollutants.

It is found that the floor-ceiling heating system has the lowest energy, exergy, CO2 emissions, operating costs, and the nominal power of the boiler. The worst system by all these parameters is the classical ceiling heating. Also, it is important to note that the next better system is system 2 (wall heating panel).

The comparison of the room air temperatures and the design temperatures shows that all systems give satisfactory results without significant deviations. The largest deviation is found with two-sided ceiling panel system 4 and the temperature in this room is 18.9 $^o$C. However, it have in mind that the using of the radiation heaters may achieve thermal comfort with up to 2°C lower temperature of internal air than that required by the design. The surrounding inner surfaces of the exterior walls have the lowest temperature at 17.5 $^o$C that occurs with the floor-ceiling panel system in room SS3 at the first floor.

## ACKNOWLEDGMENT

This paper is a result of two investigations: (1) project TR33015 of Technological Development of Republic of Serbia, and (2) project III 42006 of Integral and Interdisciplinary investigations of Republic of Serbia. The first project is titled "Investigation and development of Serbian zero-net energy house", and the second project is titled "Investigation and development of energy and ecological highly effective systems of poly-generation based on renewable energy sources. We would like to thank to the Ministry of Education and Science of Republic of Serbia for their financial support during these investigations.